# The collision between positronium (Ps) and muonium (Mu)

**Hasi Ray**[*,1,2,3]
[1]*Study Center,* S-1/407, B. P. Township, Kolkata 700094, India
[2]*Department of Physics, New Alipore College,* New Alipore, Kolkata 700053, India
[3]*National Institute of T.T.T. and Research,* Salt Lake City, Kolkata 700106, India

E-mail: hasi_ray@yahoo.com

**Abstract.** The collision between a positronium (Ps) and a muonium (Mu) is studied for the first time using the static-exchange model and considering the system as a four-center Coulomb problem in the center of mass frame. An exact analysis is made to find the s-wave elastic phase-shifts, the scattering-lengths for both singlet and triplet channels, the integrated/total elastic cross section and the quenching cross section due to ortho to para conversion of Ps and the conversion ratio.

1**. Introduction**

The importance of the studies on positron ($e^+$) and positronium (Ps) is well known [1,2]. The Ps is an exotic atom which is itself its antiatom and the lightest isotope of hydrogen (H); its binding energy is half and Bohr radius is twice that of H. The muonium (Mu) is a bound system of a positive muon ($\mu^+$) and an electron ($e^-$) and also a hydrogen-like exotic atom. Its nuclear mass is one-ninth (1/9) the mass of a proton; the ionization potential and the Bohr radius are very close to hydrogen atom. Mu was discovered by Hughes [3] in 1960 through observation of its characteristic Larmor precession in a magnetic field. Since then research on the fundamental properties of Mu has been actively pursued [4-8], as has also the study of muonium collisions in gases muonium chemistry and muonium in solids [9]. Generally chemical properties of an element are dependent on ionization potential and the Bohr radius. In Mu these are very close to H. However the vibrational frequencies involving Mu are higher than H.

We study the elastic collision between Ps and Mu when both are in ground states. The system is studied in center of mass frame using the static-exchange model (SEM) [10]. Here the Schrodinger equation of motion is solved exactly considering the system as a four-center Coulomb problem [11]. Both the direct and exchange amplitudes are calculated exactly. The H-like atomic wavefunctions are used to describe the atoms. There are total six Coulomb interaction terms in the system, from them two are used to describe the atoms. So effectively there are four Coulomb interaction terms between the two H-like atoms and there are total eight different terms in the amplitude to consider both the direct and exchange channels. In addition, there is one off-shell term for the rearrangement channel to include the effect of inelastic flux. We study the singlet (+) and the triplet (-) s-wave elastic phase-shifts ($\delta_0$) and the corresponding scattering lengths ($a\pm$) utilizing the effective range theory.

---

[*] HR, Study Center, S-1/407/6, B. P. Township, Kolkata 700094, India
.

Again the annihilation is an important phenomenon in Ps and gas atom collisions. The positrons emitted from a radioactive source into a gas at normal pressure do not suffer annihilation by collision with atomic electrons, before they have a high probability of capturing an electron from a gas atom to form Ps [12]. If the captured electron ($e^-$) has the opposite spin to the positron ($e^+$) forming para-Ps, the lifetime towards mutual annihilation is only about $10^{-10}$ sec. But if the spins are parallel, forming ortho-Ps, the lifetime is much longer about $10^{-7}$ sec. The mutual annihilation of para-Ps is associated with emission of two $\gamma$-ray photons each having energy 511 Mev. Annihilation may occur due to the exchange collision of Ps with gas atoms by which ortho-Ps transforms into para-Ps, the phenomenon is known as quenching. So it is of interests to study quenching in Ps and Mu collision when both are in ground states. We study the total elastic cross section ($\sigma$), the quenching cross section ($\sigma_q$) and the conversion ratio for Ps(1s)-Mu(1s) collision. An exact and ab-initio SEM code recently introduced by Ray [11,13] to study a four-body Coulomb problem in the center of mass frame is used.

## 2. Theory

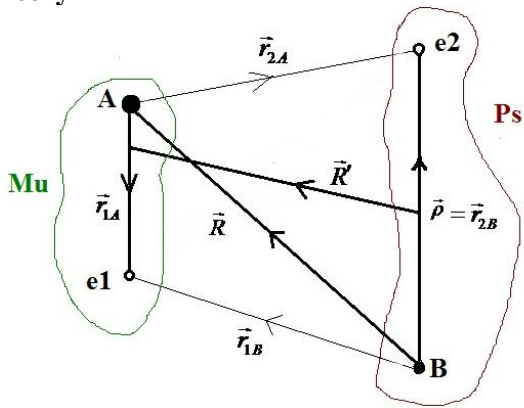

**Figure 1.** The picture of Ps and Mu system.

The description of the present Ps and Mu system is made in **Figure 1**. The $\mu^+$ of the Mu and the $e^+$ of Ps are placed at points A and B respectively; e1 and e2 are the two electrons attached to the nuclei A and B which are at a distance $r_{1A}$ and $r_{2B}$ from A and B respectively. Accordingly $\vec{r}_{2A}$ and $\vec{r}_{1B}$ are defined. $\vec{R}$ is the vector joining A and B and $\vec{R}'$ is the vector joining the center of masses of the two atoms so that

$$\vec{R}' = \vec{R} + \frac{m_e}{m_A + m_e}\vec{r}_{1A} - \frac{m_e}{m_B + m_e}\vec{r}_{2B} \qquad (2.1),$$

when $m_A$, $m_B$ and $m_e$ are the masses of $\mu^+$, $e^+$ and $e^-$ respectively. The initial and final state wavefunctions of the system are defined as

$$\psi_i = e^{i\vec{k}_i \cdot \vec{R}'} \phi^A_{1s}(\vec{r}_{1A})\eta^B_{1s}(\vec{r}_{2B}) \qquad (2.2a),$$

$$\psi_f = (1 \pm P_{12})e^{i\vec{k}_f \cdot \vec{R}_f} \phi^A_{1s}(\vec{r}_{1A})\eta^B_{1s}(\vec{r}_{2B}) \qquad (2.2b),$$

so that $\vec{k}_i$ and $\vec{k}_f$ represent the initial and final momenta of the projectile, $\phi^A_{1s}(\vec{r}_{1A})$ and $\eta^B_{1s}(\vec{r}_{2B})$ represent the ground state wavefunctions of Mu and Ps respectively and $P_{12}$ is the exchange or antisymmetry operator; $\vec{R}_f$ is defined as

$$\vec{R}_f = \vec{R} + \frac{m_e}{m_A + m_e}(\vec{r}_{2B} - \vec{R}) - \frac{m_e}{m_B + m_e}(\vec{r}_{1A} + \vec{R}) \qquad (3).$$

The Coulomb interaction between the atoms in the direct and exchange channels are defined as

$$V_{Direct}(\vec{R}, \vec{r}_{1A}, \vec{r}_{2B}) = \frac{\mu_A^+ e_B^+}{R} - \frac{\mu_A^+ e_2^-}{|\vec{R} - \vec{r}_{2B}|} - \frac{e_B^+ e_1^-}{|\vec{R} + \vec{r}_{1A}|} + \frac{e_1^- e_2^-}{|\vec{R} + \vec{r}_{1A} - \vec{r}_{2B}|} \quad (4a),$$

$$V_{Exchange}(\vec{R}, \vec{r}_{1A}, \vec{r}_{2B}) = \frac{\mu_A^+ e_B^+}{R} - \frac{\mu_A^+ e_1^-}{|\vec{r}_{1A}|} - \frac{e_B^+ e_2^-}{|\vec{r}_{2B}|} + \frac{e_1^- e_2^-}{|\vec{R} + \vec{r}_{1A} - \vec{r}_{2B}|} \quad (4b),$$

respectively. Here the magnitudes of all the Coulomb terms in the numerators in equation (4a) and (4b) are equal to unity in atomic unit (a.u.).

In elastic collision $|k_i| = |k_f|$, so only the direction of final momentum ($\vec{k}_f$) changes due to scattering or collision.

## 3. Results and Discussion

The present s-wave elastic phase shifts for the singlet ($\delta_0^+$) and triplet ($\delta_0^-$) spin-states of the two system electrons are presented in **figure** 1 and the corresponding plot of the $k^2$ versus $k\cot(\delta_0)$ are presented in **figure** 2. Following effective range theory one can easily calculate the scatteing lengths from the cut-point of the curve with y-axis. It should be noted that all the curves are almost similar to Ps and H system when both are in ground states. Both the scattering lengths are slightly greater than Ps and H system. In Ps-Mu system the singlet scattering length is 7.45 a.u. and the triplet scattering length is 2.50 a.u. whereas in Ps-H system these are 7.25 a.u. and 2.49 a.u. It is useful to note that the reduced masses of Ps-H and Ps-Mu systems are 1.9978 a.u. and 1.9807 a.u. respectively.

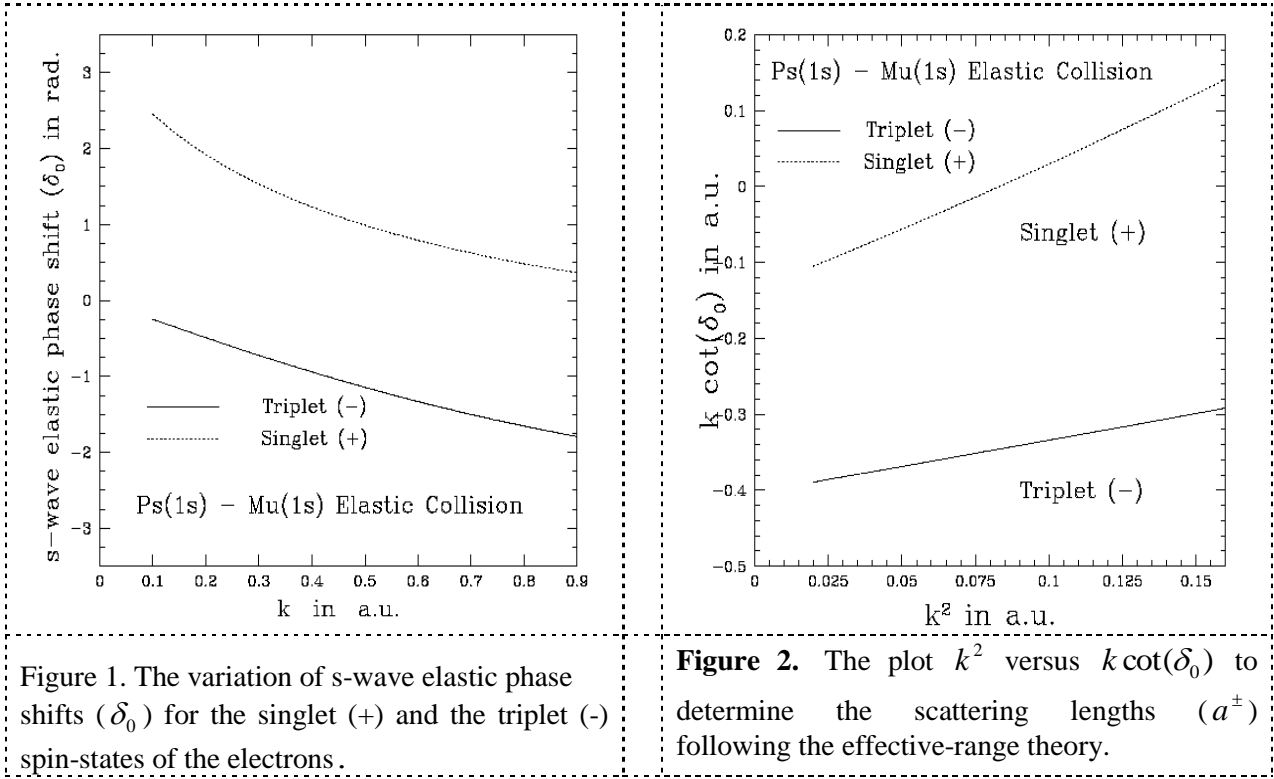

Figure 1. The variation of s-wave elastic phase shifts ($\delta_0$) for the singlet (+) and the triplet (-) spin-states of the electrons.

**Figure 2.** The plot $k^2$ versus $k\cot(\delta_0)$ to determine the scattering lengths ($a^\pm$) following the effective-range theory.

The integrated or total elastic cross section ($\sigma$) and the quenching cross section ($\sigma_q$) using the present SEM theory are presented in **figure 3**. An augmented-Born approximation is used to include the effect of higher partial waves more accurately. These data are also very close to Ps-H system. In

**figure 4**, the ortho- to para- conversion ratio ($\sigma/\sigma_q$) for Ps-Mu system are compared with Ps-H system. Both almost coincide.

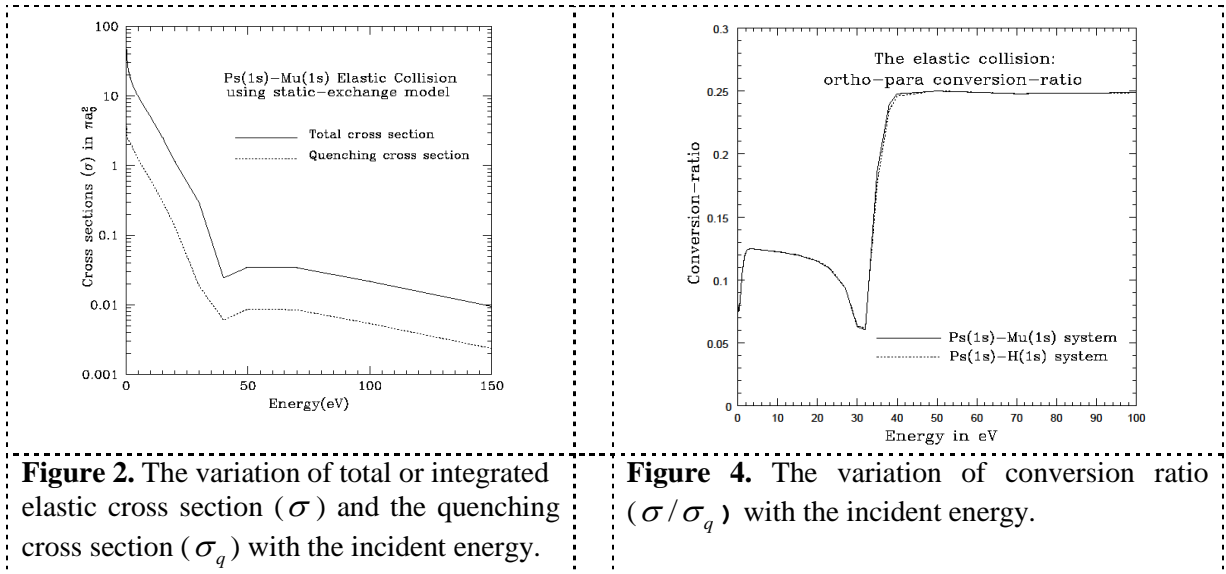

**Figure 2.** The variation of total or integrated elastic cross section ($\sigma$) and the quenching cross section ($\sigma_q$) with the incident energy.

**Figure 4.** The variation of conversion ratio ($\sigma/\sigma_q$) with the incident energy.

## 4. Conclusion

We study the Ps and Mu elastic collision using the static-exchange model (SEM) considering the system as a four center Coulomb problem in the center of mass frame. The s-wave elastic phase shifts, the scattering lengths for the singlet and triplet spin-states of the system electrons, the integrated/total elastic cross section, the quenching cross section and ortho-para conversion ratio for Ps(1s)-Mu(1s) elastic scattering are being reported for the first time.